\input{aipcheck}

\documentclass[
    ,final            
  ]
  {aipproc}

\layoutstyle{6x9}

\usepackage{amsfonts}
\usepackage{amsmath}
\usepackage{graphicx}
\usepackage{dcolumn}
\usepackage{bm}

\begin{document}

\title{Two-photon decay of a light scalar quark-antiquark state}

\classification{012.39.Ki,13.25.Jx,13.30.Eg,13.40.Hq}
\keywords      {$\sigma$-meson, relativistic quark model, electromagnetic decay}

\author{Francesco Giacosa}{
  address={Institut f\"{u}r Theoretische Physik, Johann Wolfgang Goethe-Universit\"{a}t,
Max von Laue-Str. 1, 60438 Frankfurt, Germany }
}

\begin{abstract}
The two-photon decay of a light scalar quarkonium is evaluated in a local and
a nonlocal approaches. It is shown that the two-photon decay, driven by a
triangle quark-loop diagram, is smaller than 1 keV for a mass below 0.7 GeV.

\end{abstract}

\maketitle


\section{Introduction}

It is still debated if scalar resonances below 1 GeV are quarkonia
\cite{scadron}, tetraquark \cite{tqgen}, molecular states \cite{mesonicmol1}
or a mixing of these configurations \cite{fariborz}. If, as suggested in Ref.
\cite{noqq}, the light scalar resonances are not quarkonia one is lead to
identify the $\overline{q}q$-states with resonances above 1 GeV where mixing
with the scalar glueball takes place \cite{refs}. The two-photon decays of
scalar resonances, both below and above 1 GeV, are regarded as an important
source of informations toward the understanding of this puzzle
\cite{pennington}. However, as pointed out in Ref. \cite{sgg}, care is needed
when dealing with this process.

In this work, following Refs. \cite{sgg,faessler,giacosa1}, I concentrate on
the theoretical description of the two-photon decay of a light
scalar-isoscalar quark-antiquark bound state (quarkonium). The amplitude for
this process, occurring via a triangle-loop of constituent quarks as in Fig.
1, is studied in a local and a nonlocal field theoretical approaches. In the
former case (Fig. 1.a) the scalar field, describing the quark-antiquark bound
state, interacts locally with its constituents -the quarks. In the latter case
(Fig. 1.b) a nonlocal interaction is introduced, which allows for a realistic
treatment of the finite dimension (mean radius of about 0.5 fm) of the
quark-antiquark bound state. For this reason, while interesting analytic
formulas can be derived in the local treatment, the numerical results of the
nonlocal approach represent the here outlined theoretical predictions.

In this work it is shown that the decay width of a scalar-isoscalar
quark-antiquark state, with flavor wave-function $\overline{n}n=\sqrt{\frac
{1}{2}}(\overline{u}u+\overline{d}d)$ and a mass below 0.7 GeV, is smaller
than 1 keV. This is against the common belief that a decay width of about 3 -
5 keV would favor a quarkonium interpretation of the resonance $\sigma\equiv
f_{0}(600)$. A similar study is here performed for a $\overline{s}s$ scalar
bound state: the decay into two photons via the quark-loop diagram of Fig. 1.b
is rather small, thus a dominant $\overline{s}s$ wave function of the
resonance $f_{0}(980)$ cannot explain the measured $\gamma\gamma$ decay as
reported in Ref. \cite{pdg}.

\section{Scalar quarkonium into $\gamma\gamma$}

\subsection{Amplitude in the local approach}%

\begin{figure}
\includegraphics[
height=1.7781in,
width=4.2384in
]{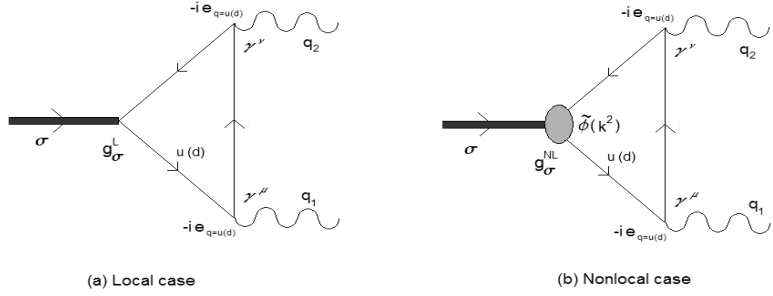}
  \caption{Diagrams for the two-photon decay of a scalar quarkonium. Left panel:
local case, in which the scalar field is point-like. Right panel: nonlocal
case, in which the scalar field has a finite dimension.}
\label{fig1}
\end{figure}

Following Ref. \cite{sgg} we consider a local interaction Lagrangian (L)
describing the (point-like, zero-radius) scalar quarkonium field $\sigma$ with
the constituent quarks $u$ and $d$:
\begin{equation}
\mathcal{L}_{\mathrm{int}}^{\text{L}}=\,\frac{g_{\sigma}^{L}}{\sqrt{2}}%
\sigma(x)\,\bar{q}(x)q(x) \label{locallag}%
\end{equation}
where $q^{T}=(u,d)$ is the quark doublet with mass $m_{q}=m_{u}=m_{d}$
(isospin limit), which can be chosen between 0.25 and 0.45 GeV as a variety of
low-energy effective approaches confirm. In this work I use $m_{q}=0.35$ GeV
(similar to the value of Ref. \cite{hatsuda}). As shown in Ref. \cite{sgg}
varying the mass in the above range changes only slightly the results. The
decay of $\sigma$ into $\gamma\gamma$ is obtained by evaluating the diagram of
Fig. 1.a:
\begin{equation}
\Gamma_{\sigma\gamma\gamma}^{\text{L}}=\frac{\pi}{4}\,\alpha^{2}M_{\sigma}%
^{3}\,\left\vert A^{\text{L}}\right\vert ^{2}\,,\text{ }A^{\text{L}}%
=\,\frac{g_{\sigma}^{\text{L}}}{2\pi^{2}}N_{c}Q_{\sigma}I_{\sigma}^{\text{L}}
\label{sgglocal}%
\end{equation}
where $\alpha$ is the fine structure constant, $N_{c}=3$ the number of colors,
the charge factor $Q_{\sigma}=\frac{1}{\sqrt{2}}(\frac{4}{9}+\frac{1}%
{9})=\frac{5}{9\sqrt{2}}$ corresponds to the flavor wave functions
$\sigma\equiv\sqrt{\frac{1}{2}}(\overline{u}u+\overline{d}d)$ and $M_{\sigma
}\,$is the mass of the scalar state. The coupling constant $g_{\sigma
}^{\text{L}}=\sqrt{2}m_{q}/F_{\pi}$ with $F_{\pi}=92.4$ MeV is obtained by
using the Goldberger-Treiman relation and linear realization of chiral
symmetry. The amplitude $I_{\sigma}^{\text{L}}$ reads \cite{sgg,faessler}:
\begin{equation}
I_{\sigma}^{\text{L}}=\frac{2m_{q}}{M_{\sigma}^{2}}\left[  1+\left(
1-\frac{4m_{q}^{2}}{M_{\sigma}^{2}}\right)  \arcsin^{2}\left(  \frac
{M_{\sigma}}{2m_{q}}\right)  \right]  . \label{isigmaloc}%
\end{equation}
Note that $I_{\sigma}^{\text{L}}$ is a real number only if $M_{\sigma}%
\leq2m_{q}$. For $M_{\sigma}>2m_{q}$ an imaginary part, due to the absence of
confinement and to the unphysical decay of the sigma meson into a
quark-antiquark pair, arises. Thus, extending the calculation to $M_{\sigma
}>2m_{q}$ is a dangerous step. Moreover, as shown later the local approach can
be trusted only if $M_{\sigma}$ is well below $2m_{q}$.

Interestingly, the $\gamma\gamma$ decay of the Higgs particle in the Standard
Model \cite{djouadi} proceeds similarly via leptonic loops as in Fig. 1.a. In
most cases $M_{\text{Higgs}}>2m_{\text{lepton}}$ and an imaginary part is
present. Being the Higgs an elementary particle interacting locally with the
leptons the use (with due changes) of Eq. (\ref{sgglocal}) is allowed below
and above threshold. This, however, is not the case for the `Higgs' of QCD,
i.e. the scalar quarkonium state, for essentially two reasons: (i)
Confinement, as mentioned above. No imaginary part shall arise. (Moreover, the
quark propagator and the coupling constants run already at low energy.) (ii)
The scalar quark-antiquark bound state is not elementary: a nonlocal
interaction with its constituents should be considered as I do in the following.

\subsection{Amplitude in the nonlocal approach}

I turn to the scalar quark-antiquark field $\sigma$ by using the following
nonlocal (NL) interaction Lagrangian \cite{sgg,faessler,giacosa1}
\begin{equation}
\mathcal{L}_{\mathrm{int}}^{\text{NL}}(x)\,=\,\frac{g_{\sigma}^{\text{NL}}%
}{\sqrt{2}}\sigma(x)\,\int d^{4}y\,\Phi(y^{2})\,\bar{q}(x+y/2)q(x-y/2)\,,
\label{lnonloc}%
\end{equation}
where the delocalization takes account of the extended nature of the
quarkonium state by the covariant vertex function $\Phi(y^{2})$. The
(Euclidean) Fourier transform of this vertex function is taken as
$\widetilde{\Phi}(k_{E}^{2})=\exp(-k_{E}^{2}/\Lambda^{2}),$ also assuring
UV-convergence of the model. The cutoff parameter $\Lambda$ can be varied
between $1$ and $2$ GeV, corresponding to a spatial extension of the $\sigma$
of about $l\sim1/\Lambda\sim0.5$ fm. Previous studies~\cite{anikin} have shown
that the precise choice of $\widetilde{\Phi}(k_{E}^{2})$ affects only slightly
the result, as long as the function falls of sufficiently fast at the energy
scale set by $\Lambda$. Moreover, the dependence of the results on $\Lambda$
is very soft \cite{sgg}. In this work the intermediate value $\Lambda=1.5$ GeV
is used for numerical evaluations.

Within the nonlocal treatment the coupling $g_{\sigma}^{\text{NL}}$ is
determined by the so-called compositeness condition $Z_{\sigma}=1-(g_{\sigma
}^{\text{NL}})^{2}\Sigma_{\sigma}^{\prime}(M_{\sigma}^{2})=0$
\cite{faessler,weinberg}, where $\Sigma_{\sigma}^{\prime}$ is the derivative
of the $\sigma$-meson mass operator given by
\begin{equation}
\Sigma_{\sigma}(p^{2})=-N_{c}\int\frac{d^{4}k}{(2\pi)^{4}i}\,\widetilde{\Phi
}^{2}(-k^{2})\,\mathrm{tr}\left[  S_{q}(k+p/2)S_{q}(k-p/2)\right]  \,,
\label{gsigma}%
\end{equation}
and $S_{q}(k)=(m_{q}-\gamma^{\mu}\!k_{\mu})^{-1}$ is the quark propagator.
Note, the compositeness condition is equivalent to the hadron wave function
normalization in quantum field approaches based on the solution of the
Bethe-Salpeter/Faddeev equation~\cite{roberts}. At this level $g_{\sigma
}^{\text{NL}}$ is a slowly decreasing function of $M_{\sigma}$ (details in
Refs. \cite{sgg,pagliara}).

Following Refs. \cite{faessler,giacosa1} the contribution of the
gauge-invariant part\footnote{Gauge invariance in a nonlocal approach implies
that other diagrams, in which the photon couples directly to the nonlocal
interaction vertex, are present \cite{faessler}. However, their contribution
is numerically suppressed of a factor $10^{-5}$ and thus is omitted here.} of
the triangle diagram of Fig. 1.b to the two-photon decay width is given by:
\begin{equation}
\Gamma_{\sigma\gamma\gamma}^{\text{NL}}=\frac{\pi}{4}\alpha^{2}M_{\sigma}%
^{3}\left\vert A^{\text{NL}}\right\vert ^{2}\,,\text{ }A^{\text{NL}}%
=\,\frac{g_{\sigma}^{\text{NL}}}{2\pi^{2}}N_{c}Q_{\sigma}I_{\sigma}%
^{\text{NL}} \label{sggnonlocal}%
\end{equation}
with $I_{\sigma}^{\text{NL}}=I_{\sigma}^{(1)}+I_{\sigma}^{(2)}\,$and
\begin{align}
I_{\sigma}^{(1)}  &  =m_{q}\int\frac{d^{4}k}{\pi^{2}i}\,\widetilde{\Phi
}(-q^{2})\,\frac{1}{(m_{q}^{2}-p_{1}^{2})(m_{q}^{2}-p_{2}^{2})(m_{q}^{2}%
-p_{3}^{2})}\,,\label{int1}\\
I_{\sigma}^{(2)}  &  =-m_{q}\int\frac{d^{4}k}{\pi^{2}i}\,\widetilde{\Phi
}(-q^{2})\,\frac{{\frac{4}{M_{\sigma}^{2}}k^{2}-\frac{32}{M_{\sigma}^{4}}%
}(kq_{1})(kq_{2})}{(m_{q}^{2}-p_{1}^{2})(m_{q}^{2}-p_{2}^{2})(m_{q}^{2}%
-p_{3}^{2})}\,. \label{int2}%
\end{align}
where $q_{1}$ and $q_{2}$ are the photon momenta and $p_{1}=k+q_{1},$
$p_{2}=k,$ $p_{3}=k-q_{2},$ $q=(p_{1}+p_{3})/2.$ Clearly the limit
$\lim_{\Lambda\rightarrow\infty}I_{\sigma}^{\text{NL}}=I_{\sigma}^{\text{L}}$ holds.

\section{Results and discussions}

We report in Fig. 2 the amplitude and the two-photon decay width in both the
local and nonlocal theories as function of the mass $M_{\sigma}.$ A number of
comments is in order:%

\begin{figure}
\includegraphics[
height=2.2295in,
width=4.5593in
]{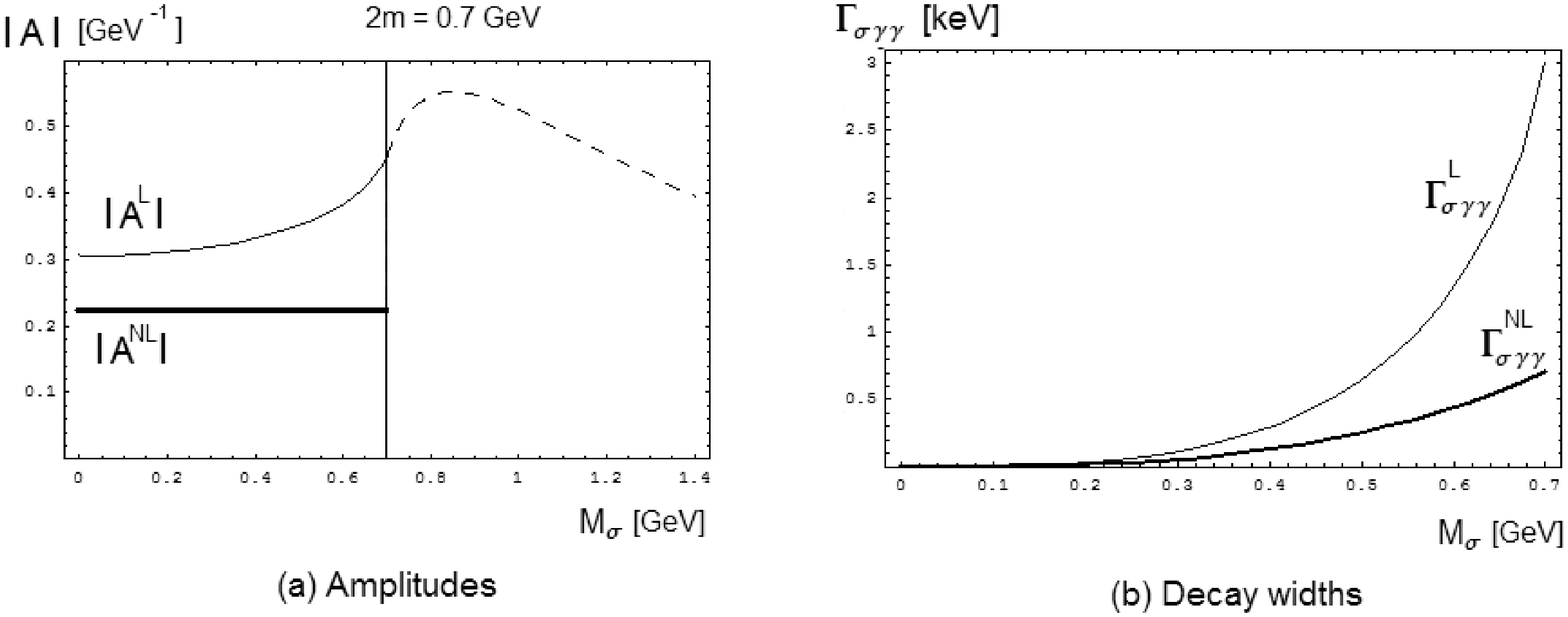}
\caption{Left panel: $\gamma\gamma$-amplitude in the local (thin solid line
below $2m_{q}$ and dashed line above $2m_{q}$) and in the nonlocal (thick
line) cases. Right panel: decay width within the local (thin line) and
nonlocal (thick line) approaches. The values $m_{q}=0.35$ GeV and
$\Lambda=1.5$ GeV have been used.}
\label{fig2}
\end{figure}

(a) The nonlocal amplitude $\left\vert A^{\text{NL}}\right\vert $ of Eq.
(\ref{sggnonlocal}) (thick line in Fig. 2.a) shows an almost constant behavior
from $0$ up to threshold. This is a remarkable and stable result. Notice that
$\left\vert A^{\text{NL}}/A^{\text{L}}\right\vert =0.73$ for $M_{\sigma
}\rightarrow0$ because of the cutoff $\Lambda$ (i.e., finite dimension) in
Eqs. (\ref{int1})-(\ref{int2}). Contrary to the nonlocal case, the local
amplitude $\left\vert A^{\text{L}}\right\vert $ of Eq. (\ref{sgglocal}) (solid
thin line in Fig. 2.a) is enhanced at threshold. This fact is however
connected to the non-realistic point-like nature of the sigma field and to
the, also non-realistic, constant (not running) behavior of the coupling
strength $g_{\sigma}^{\text{L}}.$ In fact, while $g_{\sigma}^{\text{NL}%
}(M_{\sigma}\rightarrow0)=5.56$ $\simeq g_{\sigma}^{\text{L}}=\sqrt{2}%
m_{q}/F_{\pi}=5.36$, one has just below threshold $g_{\sigma}^{\text{NL}%
}(M_{\sigma}=0.69)=3.48,$ thus sizably reduced.

(b) The dashed line in Fig. 2.a describes $\left\vert A^{\text{L}}\right\vert
$ above threshold, see Eq. (\ref{isigmaloc}). As remarked previously the local
amplitude $A^{\text{L}}$ is a complex number above $2m_{q}$; this is a
consequence of lack of confinement. The nonlocal amplitude $\left\vert
A^{\text{NL}}\right\vert $ has been plotted only up to threshold. Although it
would be appealing to continue the constant behavior even above $2m_{q}$, this
cannot be done at the present level. A consistent modification of the quark
propagator should be applied to study the reaction above $2m_{q}.$

(c) In Fig. 2.b the decay width as function of $M_{\sigma}$ is shown. While
the local and nonlocal approaches deliver similar results for small
$M_{\sigma}$ large differences arise for $M_{\sigma}\rightarrow2m_{q},$ where
$\Gamma_{\sigma\gamma\gamma}^{\text{L}}$ overshoots $\Gamma_{\sigma
\gamma\gamma}^{\text{NL}}$ of a factor 4.

Thus, the final result of the present study is summarized in the following
equation:
\begin{equation}
\Gamma_{\sigma\gamma\gamma}^{\text{NL}}<1\text{ keV for }M_{\sigma}<0.7\text{
GeV.}%
\end{equation}
In Ref. \cite{sgg} a more detailed study on parameter variations confirms this
result. In Ref. \cite{scadron} the amplitude is calculated within the local
treatment in the limit $M_{\sigma}\rightarrow2m_{q}$ and thus a decay widths
larger than 1 keV is obtained\footnote{Note also that a small change in
$M_{\sigma}$ causes a large change of $\Gamma_{\sigma\gamma\gamma}^{\text{L}}$
(Fig. 2.b) thus making a clear prediction close to threshold difficult.}.
However, for all the reasons discussed above I would like to point out that
the local approach should not be used for $M_{\sigma}\rightarrow2m_{q}$ where
the discrepancy with the nonlocal treatment is most evident. Above threshold
the application of the local approach is even more problematic. I conclude
this discussion by stressing that the region of applicability of the local
approach is restricted to small $M_{\sigma}$ (safely below $2m_{q}$).

As a last step I evaluate the $\gamma\gamma$ decay width of a scalar state
$S\equiv\overline{s}s$ within the nonlocal approach. The corresponding
$\Gamma_{S\gamma\gamma}^{\text{NL}}$ has the same form of Eq.
(\ref{sggnonlocal}) up to the charge factor which is now $Q_{S}=1/9.$ By using
a constituent strange mass $m_{s}=550$ MeV (close to the value of Ref.
\cite{hatsuda}) and $\Lambda=1.5$ GeV a decay width $\Gamma_{S\gamma\gamma
}^{\text{NL}}=0.062$ KeV is obtained. Thus, a dominant $\overline{s}s$
component of the resonance $f_{0}(980)$ cannot explain -via the quark-loop of
Fig. 1.b - the experimental value $\Gamma_{f_{0}(980)\rightarrow\gamma\gamma
}=0.39_{-0.13}^{+0.10}$ \cite{pdg}. The reason why in Ref. \cite{scadronss} a
larger value of 0.3 keV is obtained has essentially the same origin (local
treatment at threshold) explained in the non-strange case.

\section{Conclusions}

The $\gamma\gamma$ transition of a scalar quarkonium has been studied in a
local and a nonlocal field theoretical approaches. The limit of validity of
the local approach has been carefully addressed. Within the nonlocal
treatment, which takes into account the finite dimension of the quarkonium
state, it is shown that the quark-loop contribution to the decay of a scalar
quarkonium with wave function $\sqrt{\frac{1}{2}}(\overline{u}u+\overline
{d}d)$ into two-photons is smaller than 1 keV for a mass below 0.7 GeV. Thus,
an eventual confirmation of a large two-photon width of the resonance
$f_{0}(600)$ does not favor a quarkonium interpretation of the latter. A
similar consideration holds for an $\overline{s}s$ interpretation of the
resonance $f_{0}(980)$. Loops of kaons and pions should be included as a
further contribution to $\gamma\gamma$ decay width which, although suppressed
according to large-N$_{c}$ counting rules, may play an important role. The
corresponding evaluation within the nonlocal treatment represent an
interesting outlook.


\begin{theacknowledgments}
I thank Th. Gutsche and V. Lyubovitskij with whom
Ref. \cite{sgg} was written.
\end{theacknowledgments}





\bibliography{sample}


\end{document}